# Observation of Rb Two-Photon Absorption Directly Excited by an Erbium-Fiber-Laser-Based Optical Frequency Comb via Spectral Control

Jiutao Wu[1], Dong Hou[1], Xiaoliang Dai[2], Zhengyu Qin[2], Zhigang Zhang[1], Jianye Zhao[1]*

[1] *Department of electronics, Peking University, Beijing, China.*
[2] *School of Physics, Peking University, Beijing, China.*

**Abstract**: We demonstrated the observation of Rb two-photon absorption directly excited by an optical frequency comb at fiber communication bands. A chain of comb spectral control is elaborately implemented to increase the power of the second harmonic optical frequency comb generation and the two-photon transition strength. A two-photon transition spectrum is obtained with clearly resolved transition lines. It provides a potential approach to realize the optical frequency comb or optical clock at ~1.5$\mu$m with high stability and accuracy.

Two-Photon transitions (TPT) of alkali metal vapors are widely used as important tools for spectroscopy and nonlinear optics in atomic physics. The Rubidium (Rb) 5S-5D TPTs have a considerable narrow linewidth (~500 kHz) and low sensitivity to external exciting lights [1]. Therefore, the TPTs lines offer a powerful way to stabilize the optical frequency and act as frequency standards with high stability. The optical frequency standards based on the Rb TPT have been established, and their performances in tern of accuracy and stability were reported to the ~$10^{-12}$ and ~$10^{-13}$ respectively [2]. The importance of these transitions were recognized in 1997, when the (Fg=3) to (Fg=5) hyperfine component of the $5^2S_{1/2}$ to $5^2S_{5/2}$ transition is recommended by the *Comit´ e International des Poids et Mesures* (CIPM) as the definition of the meter [3]. In early researches, most of Rb TPTs were excited by continuous wave (CW) laser, e.g. dye lasers or diode lasers [4, 5]. Compared to the CW lasers, fs pulses laser can perform a higher resolution spectroscopy of TPTs [6]. Benefitting from the board spectrum span and separate equal interval modes of optical frequency comb (OFC), excitation of TPT by OFC laser has the advantage to transfer the stability of the Rb two-photon at 778 nm to the optical mode of the comb with other wavelength. Furthermore, it also provides a powerful tool for quantum coherence control of the interaction between atoms and laser.

Optical frequency standard with optical communication band at ~1.5 $\mu$m is required for next generation dense wavelength-division-multiplexed (DWDM) system, precision wavelength measurement and optical spectrum analysis. To obtain the optical frequency standards in this band, two transitions stabilization schemes can be considered. One is stabilizing the laser to the atomic or molecular potential transitions in the ~1550nm wavelength range, e.g. ammonia, acetylene, and hydrogen iodide [7,8,9]. The other scheme is stabilizing the second harmonic (SH) of the laser at ~ 1.5 $\mu$m to the Rb transitions near 780 nm. In the early related Rb transitions stabilization experiments with SH generator, the bulk external SH generation in $KNbO_3$ can only produce rather low power SH signal on the order of pW [10], which was not sufficient to directly excite the Rb transitions. Therefore, another ~780 nm laser with enough power which is first stabilized to the Rb transitions, was needed for indirectly locking of the SH of the 1550 nm lasers to the transitions near 780 nm [11]. Recently, it has attracted a lot of interest to excide the Rb TPT directly by the SH of lasers at 1550 nm. The techique of quasi-phase-matched (QPM) benefits the power growth in the power of SH generation. With this technique, Rb TPT was directly excited by a frequency doubled external-cavity diode laser [12]. To our knowledge, there has been no report on directly excitation of by fs pulse laser at optical communication bands. In this paper, we firstly observe the two-photon of Rb two-photon absorption directly excited by an OFC with optical communication wavelength via a chain of spectral control. Our work provides a novel and potential approach to generate a Rb two-photon-transition-based OFC with high stability and accuracy at optical communication bands.

In our experiment, an Erbium-fiber-laser-based OFC with fs pulse laser is used as the two-photon excitation laser source. Figure 1 shows the schematic of the Rb TPT directly excited by our OFC. The relevant energy level is also shown in Fig.1. For both isotope of rubidium ($^{85}$Rb and $^{87}$Rb), the 5$s$ to 5$d$ TPTs are excited by the frequency doubled OFC and the transition strength is monitored by detecting the 420 nm fluorescence derived from the cascade 5D-6P-5S transitions. Compared to normally used Ti-sapphire laser OFC, the Erbium-fiber-laser-based OFC with center wavelength of 1560 nm has a relatively low laser power (<80 mW) in our experiment. In order to obtain sufficient TPT intensity for observation, a chain of OFC spectral control is well designed to optimize the amplitude and phase of the comb spectrum. By using this spectral control chain, we can obtain improved the SH generation and hence a relatively enhanced Doppler free TPT signals. Furthermore, the chain also processes the quantum coherence control to the TPT, eliminating the Doppler broaden background to a relatively low level.

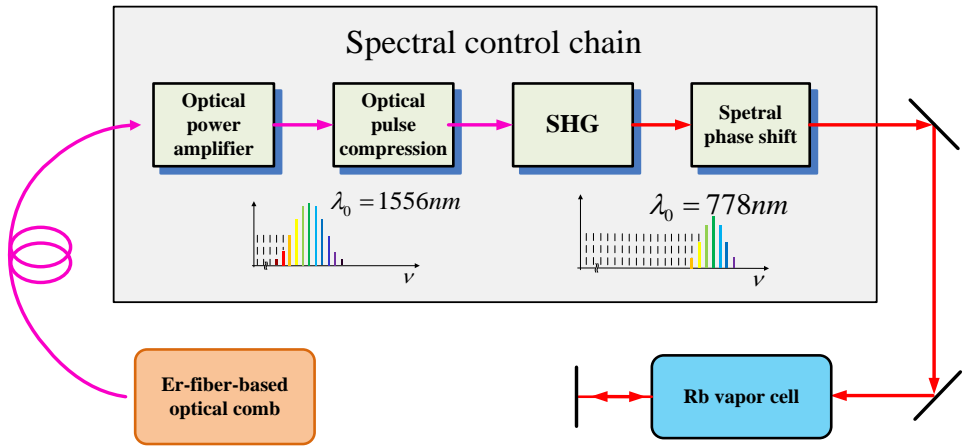

Fig. 1 The schematic of the Rb TPT directly excited by our OFC

As illustrated in Fig.1, the spectral control chain consists of four parts, which are optical power amplifier, optical pulse compression, second harmonic generation (SHG) and spectral phase shift. In the first part of the chain, an erbium-doped fiber amplifier (EDFA) operates as the optical power amplifier. The amplified optical fs pulse beam is transferred to the optical pulse compression part. In the second stage, we use a pair of silicon prisms to compensate the dispersion between OFC modes and thus obtain transform limited pulses with high peak power. In the third stage, a MgO doped periodically poled lithium niobate (PPLN) waveguide frequency doubler is used for SH generation of the OFC. The SH generation efficiency is proportional to the square of electric field intensity of the input fundamental laser, and the optical pulse duration and peak electric field power have great influence on the SH generation efficiency. Therefore the performances of last two parts will directly determine the SH generation efficiency. After the third part, the optical frequency comb is frequency doubled and transferred to an optical frequency comb with center wavelength of ~778 nm. TPTs excided by OFC brings a problem that Doppler free TPT only happens in the interaction region of the two counter-propagating pulses, thus its signals are easily obscured by the Doppler broaden signals. To overcome the problem, in the fourth part, a quantum coherence control technique, which introduces a time delay between two parts of spectral components that are above and below the resonant frequency respectively, is used to eliminate the Doppler broaden ground signals. Benefitting from this fourth part of the frequency control chain, clear and obvious Rb TPTs signals were observed with eliminated Doppler broaden background.

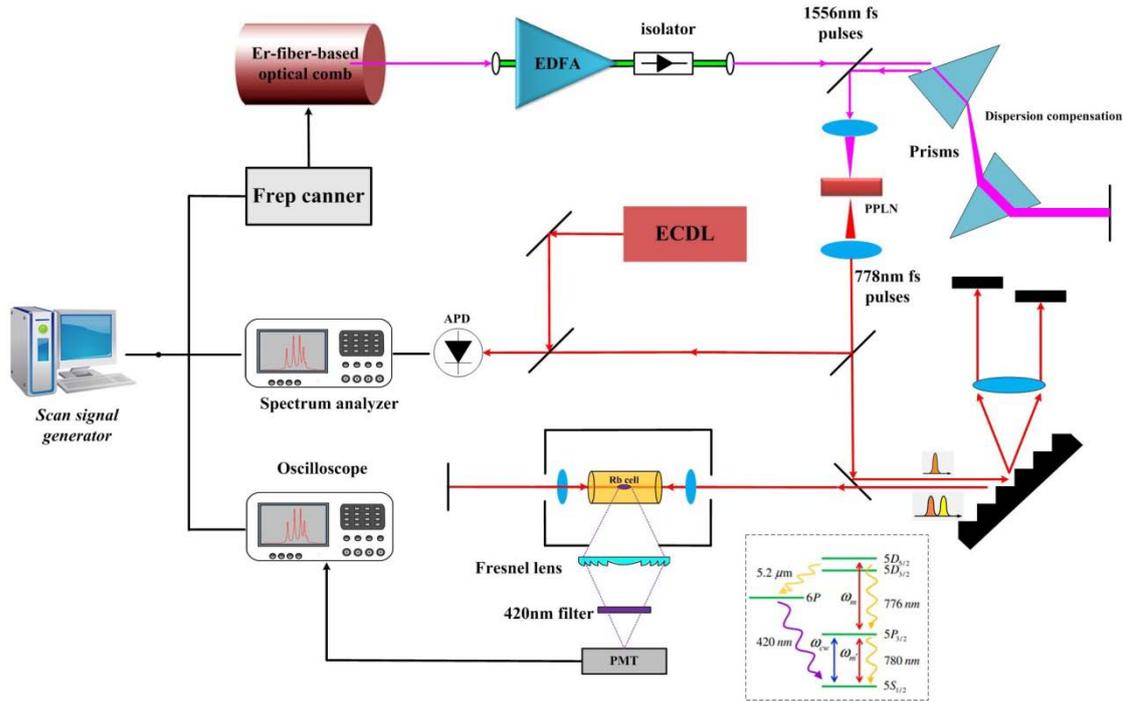

Fig. 2 Experiment setup of Rb TPT directly excited by our OFC

A diagram of the experimental apparatus is shown in Fig.2. The laser source is an Erbium-doped-laser-based OFC that is capable of generating optical pulses from 1540 nm to 1580 nm with 50 mW average power and a repetition rate ($f_{rep}$) of 144.5MHz.The absolute frequency of each OFC mode can be detuned by slightly changing the $f_{rep}$, which is realized by adjusting the laser cavity length via a PZT. The OFC power is amplified to ~100 mW by an EDFA. In order to get transform limited pulse with high peak power and hence improved SH generation efficiency, a couple of prisms is used to compensate the OFC dispersion. The prisms are made of silica for it has large dispersion index at 1.5 $\mu m$ communication bands. By control the light path between the two prisms, the dispersion is compensated, leading to transform limited pulses with duration of ~70fs. The optimized OFC laser beam is then focused into the PPLN waveguide by a lens with focal length f=30 mm for optimum efficiency generation of SH OFC. The PPLN waveguide has nine separate poled gratings with a length of 1 mm and poling periods ranging from 18.5 $\mu m$ to 20.9 $\mu m$ A 20 mW SH OFC with 10 nm optical spectrum range, centered at 778nm, is obtained with the 19.4 $\mu m$ poled grating whose poling period is precisely adjusted via temperature control to match the wavelength of transitions.

The power level achieved by this SH generation process is adequate for Rb TPTs. However, as mentioned above, Doppler free TPT signals are apt to be obscured by strong Doppler-broadened background. The Doppler-free signals derive from the TPTs of Rb atoms which absorb two photons simultaneously from oppositely directed laser beams, while transitions excided by two co-propagating photons result in a Doppler-broaden background. When excited by a single frequency CW laser in resonance, both Doppler-free and Doppler-broaden transitions take place along the whole laser path in Rb vapor cell. However, the fs pulses excited Doppler-free TPTs can only take place in the confined region where counter-propagating pulses overlap, with a length of several tens of $\mu$m. In this case, the Doppler-free signals are relatively weak compared with the whole path Doppler-broaden signals, which would reduce the signal to noise ratio of the narrow absorption lines. Several methods have been proposed to solve this problem, including the use of collimated atomic beams [13] and laser cooling technique [14]. However, the complexity of these approaches is a problem in the realization of experiment setup. Quantum coherence control provides an alternate way to suppress the co-propagating laser induced TPTs [15]. In our experiment, we utilize a 1200 grooves/mm grating and two mirrors to achieve the quantum coherence control. The OFC is spectral dispersed by the grating, and then collimated by a 20cm focal-length lens. Two mirrors are placed nearby and parallel to the Fourier plane with a distance of several mm. The upper 778 nm part and lower 778 nm part of the OFC spectrum are reflected by the two mirrors respectively. The distance between the mirrors introduce a time delay between the two spectrum parts, which corresponds to a phase shift of optical spectrum. The two OFC parts then retrace the path and are focused by the lens and combined again after the grating.

The shaped OFC pulses are then directed through an Rb vapor cell containing natural Rb isotopes (72% $^{85}$Rb, 28% $^{87}$Rb). After passing through the cell, the OFC laser beam is reflected by a mirror which is placed at a distance of c/2 $f_{\text{rep}}$ from the cell center, which ensures that each returning pulse overlap temporally with the following pulse in the middle of the Rb cell. The cell is placed in an oven and its temperature is stabilized around 70$^\text{o}$C, corresponding to a Rb pressure of 1.5×10$^{-3}$ Pa. The excited atoms return to the ground state through the cascade 5D–6P–5S transition and emit blue fluorescence at 420 nm. The fluorescence is collected by a Fresnel lens located on the side of the cell and then passes a 420 nm filter to avoid noise from background light and stray OFC light. A photomultiplier tube (PMT) is used for the detection of the fluorescence. The Rb cell and PMT are surrounded by a magnetic shield in order to reduce the influence of external magnetic fields.

To obtain the Rb TPT absorption spectrum excited by the frequency doubled OFC, we scan the optical frequency of the OFC and record the fluorescence signals simultaneously. The OFC frequency scanning is achieved by applying a linear voltage ramp to the laser cavity PZT, which can control the $f_{\text{rep}}$ of the OFC laser pulses precisely. The fluorescence signals from the PMT are recorded by an oscilloscope and sent to a personal computer. To get the absolute frequency of each mode of the OFC, we utilize a narrow linewidth external-cavity diode laser (ECDL) at 780 nm to beat with the frequency doubled OFC. Both the resultant beat note frequency ($f_{\text{beat}}$) and $f_{\text{rep}}$ are detected by an avalanche photodiode (APD) and sent to PC for calculate absolute frequency of each scanning mode of the OFC. The absolute frequency ($f_{abs}$) of $n$th OFC mode relatively to the ECDL frequency ($f_0$) is determined by

$$f_{abs}=f_0+f_{beat}+nf_{rep}.$$

With simultaneously recorded $f_{abs}$ and fluorescence strength, we obtained the Rb TPT absorption curve in a single scan process, as illustrated in Fig.3. Each point of the curve is the average data within 10 kHz around it.

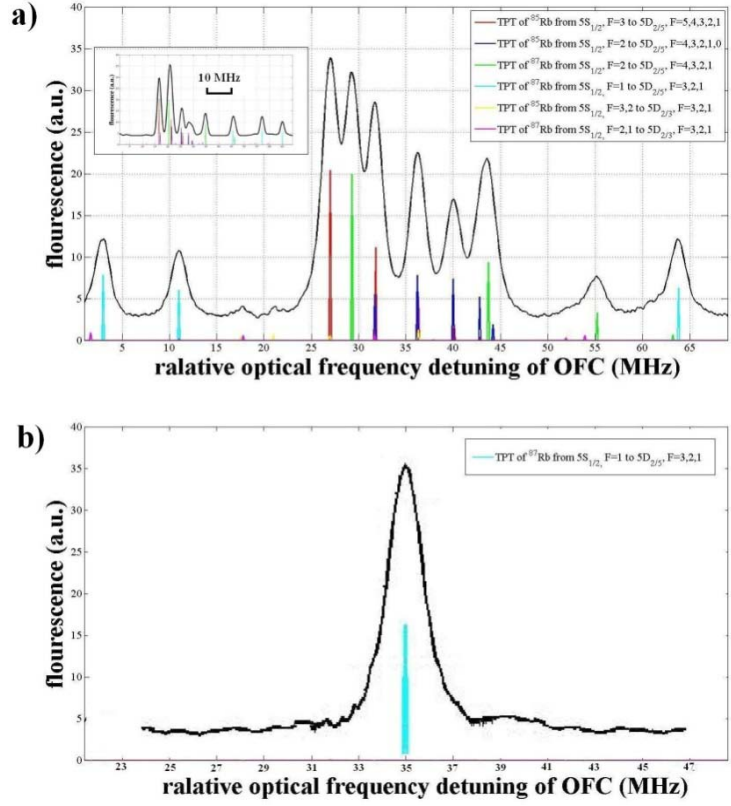

Fig. 3 Rb TPT spectrum excited directly by frequency comb

The Doppler free TPT take place when the frequency of either one OFC mode or the center of two adjacent OFC modes is equal to the TPT frequency. As a result, the TPT spectrum repeats its pattern once the OFC spectrum sweeps for half the $f_{rep}$. Hence, all the possible TPT lines appear in the scanning region with a spectral width of $f_{rep}/2$. In Fig. 3a, the $f_{rep}$ is 144.563 MHz, which correspond to 1/21 of the split of $^{85}$Rb $5S_{1/2}$ (F=2) and $5S_{1/2}$, (F=3) energy levels. In this case, four of the TPT lines of $^{85}$Rb from $5S_{1/2}$ (F=3) are overlapped with the ones from $5S_{1/2}$ (F=2). Compared to the $5S_{1/2} \rightarrow 5D_{3/2}$ transitions, the $5S_{1/2} \rightarrow 5D_{5/2}$ ones have closer proximity of the intermediate, leading to a 20 times stronger strength of TPT. In our experiment, all the $5S_{1/2} \rightarrow 5D_{5/2}$ TPT lines of both $^{85}$Rb and $^{87}$Rb can be clearly observed. For the repeating characteristic of the spectrum, some TPT lines locate nearby each other and their absorption curves are partly overlapped. By adjusting the $f_{rep}$ of the OFC laser source with a step of ~MHz, we can obtain various TPT spectral patterns, in which the positions and relatively distances of TPT peaks are different. In this way, the complete curve for each TPT absorption line can be obtained. One of the absorption lines, TPT from 87Rb $5S_{1/2}$ (F=1)→$5D_{5/2}$ (F=1) is shown in fig.3a. As we see in the figure, the Doppler broaden background stays in a relatively low level with respect to the Doppler free signals peak. The observed linewidth of this TPT transition is ~2 MHz, which is bigger than natural linewidth. Residual Doppler broadening is responsible to this broadening, In the process of TPT excided by OFC mode pairs with different frequencies, the Doppler frequency shifts of the two modes are not strictly equal, and as a result, the sum Doppler frequency shift can't cancel completely.

In conclusion, we have demonstrated TPT absorption directly excited by an OFC at the communication bands via an elaborate frequency control chain. By optimizing the amplitude and phase of the OFC laser source, we obtain an OFC with power of 20 mW and spectral span of 10 nm, centered at 778 nm. A method of quantum coherence control which introduces phase shift between frequency doubled OFC modes suppresses the Doppler broaden background to a relatively low level, resulting in a Rb TPT absorption spectrum of clearly resolved transition lines with ~2MHz linewidth. Future potential approaches to obtain a high-stability and high-accuracy OFC at fiber communication band and an optical clock can both be achieved by locking the optical frequency of the comb to the one of detected TPT.